\documentclass[journal,transmag]{IEEEtran}
\usepackage{graphicx}
\usepackage{graphics}
\usepackage{subfigure}
\usepackage{calligra}
\ifCLASSINFOpdf
\else
\fi
\begin{document}
%
\title{Time from quantum state complexity and the pace of time flow}



%
%

\author{\IEEEauthorblockN{Xiao Dong, Ling Zhou}
\IEEEauthorblockA{Faculty of Computer Science and Engineering, Southeast University, Nanjing, China}}

%



\IEEEtitleabstractindextext{%
\begin{abstract}
Based on the hypothesis that the thermodynamic arrow of time is an emergent phenomenon of quantum state complexity evolution, we further propose that the natural pace of time flow is proportional to the changing rate of quantum state complexity. we then testify how the pace of time flow changes under both special and general relativity based on the analogy between qubit quantum operations and Lorentz transformations. Our simulation results show a qualitative consistency between our hypothesis and the time dilation effect of relativity. We also checked the relationship between our idea on time with the thermal time hypothesis and we showed that our idea can be regarded as a natural generalization of the thermal time hypothesis.

\end{abstract}

}

\maketitle



\IEEEdisplaynontitleabstractindextext

%
\IEEEpeerreviewmaketitle

\section{Motivation}
The origin of spacetime has been discussed for many years. Since the last decade, the quantum information perspective began to play a role on this problem. From Lloyd's computational universe \cite{Lloyd2006A} to van Raamsdonk's building up spacetime with entanglement\cite{Raamsdonk2010Building}, ER=EPR\cite{Maldacena_ER_EPR},  Preskill's spacetime as a quantum error correcting code\cite{Pastawski2015Holographic}, machining learning geometry from entanglement features\cite{You2017Machine} and gravity from bulk entanglement\cite{Cao2017Space},  concepts in quantum information, such as entanglement, mutual information, complexity, are getting more and more related with the structure of spacetime. Along this emergent spacetime route, most works focus on the spatial component of the spacetime. How time emerges from quantum states was less addressed.

In this work we aim to develop and testify our hypothesis on the relationship between quantum state complexity and time proposed in \cite{Dong_time}, where based on a study of two simple quantum systems, a two-qubit and a three-qubit system in which a reversal of the thermodynamic arrow of time was observed, we proposed that
\begin{itemize}
  \item H1: Our observed universe as a gigantic many-body quantum system is in a low complexity state, whose quantum state complexity is linearly increasing with its evolution.
  \item H2: The observed smooth flow of time origins from the linear pattern of the state complexity of the universe.
  \item H3: The thermodynamic arrow of time points at the direction of increasing quantum state complexity.
\end{itemize}

We verified hypothesis H3 in \cite{Dong_time} by showing the consistency between the arrow of time and the change of state complexity on the above mentioned two-qubit and three-qubit systems. But we leave H1 and H2 untouched since we borrow them from Susskind's work.

In this paper we further develop our idea and propose a new hypothesis as
\begin{itemize}
  \item H4: The natural pace of time is proportional to the changing rate of the quantum state complexity.
\end{itemize}

This is to say, when the quantum state complexity changes faster with respect to a latent time variable, our clock time will run faster and vice versa. Obviously, to verify this hypothesis we need at least to show this hypothesis is consistent with the time dilation effect in both special and general relativity.

In this paper we will firstly give more evidence on the rationality of H1 and H2. Then we show that H4 is qualitatively consistent with the well known time dilation effect of a boosted inertial reference coordinate and an observer in a gravitational field.

The remaining part of this paper is arranged as follows. In section 2, we will give a detailed description of our hypotheses H1-H4. The analysis of time dilation in special and general relativity is addressed in Section 3 and 4 respectively. In section 5 we examine the relationship between our time from quantum state complexity idea with the thermal time hypothesis. Finally we give our conclusional remarks.

\section{Problem setup}

The main goal of this paper is to testify H4, which is built on H1-H3. We believe before checking H4, it's better to give a detailed discussion on H1-H3 to make H4 more concrete.

\subsection{Hypotheses H1-H3}
Our hypotheses H1-H2 are from the work of Nielsen and Susskind. In \cite{Susskind_ER_bridge_nowhere} Susskind gave a detailed discussion on the quantum state complexity based on Nielsen's geometrical picture of quantum computational complexity. His \emph{entanglement is not enough} showed that entropy and entanglement are not enough to fully describe a quantum system. Instead we have to take quantum state complexity into considerations since quantum state complexity corresponds to both spatial volume and action. In \cite{Susskind2016The} he addressed the typical state paradox of black holes. There he give a qualitative description of different phases of black holes from a computational complexity perspective as follows (a direct quote).

\emph{The ensemble of typical states is time-reversal symmetric with white holes being as likely as black holes.By contrast, natural black holes are objects that start out with low complexity, and evolve for a very long time before complexity reaches its maximum. They are black holes as long as the complexity increases, after which they have a component of white hole behavior.}

\emph{The expected behavior is that complexity increases linearly for an exponential length of time. This is the true black hole period. At the other extreme, after a doubly exponential quantum recurrence time 4.2, the complexity will become small again. Decreasing complexity defines the white hole period, during which the horizon is very vulnerable to instabilities.}

\emph{In between black and white hole behavior the system spends a time $\exp(\exp(S))$ in complexity-equilibrium. During this “grey hole” period there is no arrow of time to stabilize the geometry.}

\begin{figure}
  \centering
  \includegraphics[width=9cm]{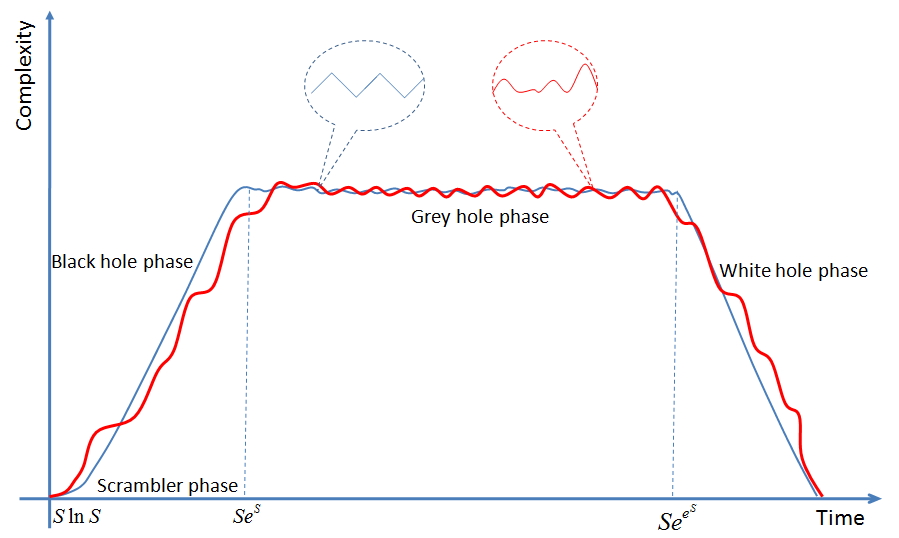}
  \caption{Different phases of black holes. This can also be used to describe the quantum state of our universe. The properties of the four phases may be different depending on the initial states and Hamiltonians as shown by the blue and red curves，which show different quantum state complexity patterns along their evolutions. Our universe may fall in the black hole phase so that we have a omnidirectional time flow but the pace of time may not be constant. We might also  in the grey hole phase and vibrate along the zig-zag of the complexity pattern.}\label{fig1}
\end{figure}

Obviously Susskind already had a clear idea on the relationship between quantum state complexity and the arrow of time. He divided the possible states of a black hole into four phases according to different patterns of quantum state complexity as shown in Fig. \ref{fig1}. They are

\begin{itemize}
  \item Scrambler phase: This is the phase starting from a simple product state where the quantum state complexity increases exponentially. This phase last for a logarithmic short time. At the end of this phase, the state complexity is roughly $Slog(S)$ with $S$ the entropy of the system.
  \item Black hole phase: This corresponds to a linear increasement of quantum state complexity lasting for an exponentially long time till the state complexity reach $exp(S)$. In this phase we have a clear arrow of time.
  \item Grey hole phase: This is a double-exponentially long time phase where the state complexity fluctuates around the maximal value $exp(S)$ and no obvious arrow of time exists.
  \item White hole phase: This is a complexity decreasing phase which is unstable and the arrow of time is reversed.
\end{itemize}

Our observed universe as a many-body quantum system under a random evolution seems to also follow the four phases. Obviously H1 and H2 correspond to the black hole phase. To complete the picture, we would like to have a detailed discussion on different phases.

The scrambler phase starts to build correlation and entanglement from a product state. The exponential expanding state complexity, which equals to the spatial volume according to Susskind, reminds us of the inflation process. Are they somehow related? A check on van Raamsdonk's idea that geometric distance is represented by mutual information\cite{Raamsdonk2010Building}\cite{Qi2013Exact} seems to deny this. Because in this phase the mutual information is built up among subsystems starting from a product state, but an increasing mutual information means a decreasing distance. How this can lead to an inflation? Do we have a contradiction here? Maybe not. Since in the initial product state all the subsystems have an infinite distance between any subsystems due to their minimal mutual entropy. With the scrambling operation, entanglement is generated so that all the subsystems are glued together by entanglement, and at the same time, the increasing mutual information means the initially separated subsystems are pulled near to each other as a result of gluing subsystems. So the \emph{inflation} here is not \emph{an expansion from a single seed}, instead it's more to pull together multiple seeds and merge them into a complete spatial space. This is different from the picture illustrated in \cite{VISHNU2010ON}, where the cosmological singularity is regarded as stemming from an unbounded high curvature manifold with a vanishing geodesic distance between any points. The very high curvatures mean a non-equilibrium condition of the dynamical instability and with the Big Bang, non-zero geodesic distance metrics appear and the evolution generates spacetime. It's worthy to point out, \cite{VISHNU2010ON} also indicated the close relationship between geodesic distance and time, which is exactly the same as our idea since the quantum state complexity is indeed a distance metric. So two different pictures of the very early stage of our universe lead to almost the same idea of time. But in \cite{VISHNU2010ON} they did not connect their geodesic distance with the concept of quantum state complexity, which is the key contribution of this paper. So the scrambler phase corresponds the early state of our universe and our universe is now not likely in this phase. Recently a similar idea was addressed in \cite{Bao2017Quantum} where the early time of the universe is regarded as a quantum circuit entangling originally unentangled ancilla qubits. But why did the \emph{universal quantum circuit} stop entangling more qubits after the inflation? Or is our expanding universe still keeping swallowing more qubits today? If our observed universe is only a subsystem of a much larger system, then the possible continuous entangling more qubits may dilute the mutual information between subsystems in the observed universe since they are entangled with more and more qubits outside the observed universe. The diluted mutual information may then lead to an expansion of the universe due to the \emph{distance from mutual information} idea.

The linear complexity increasing phase is a good candidate for the state of our universe. If we are really in this phase, then we will possess a constant flow of time and a stable thermodynamic arrow of time as we observed. But we need to be more careful. The have to ask, does the pattern complexity pattern depend on the initial state and the Hamiltonian? How can the linear complexity increasement be kept for an exponentially long time till the complexity reaches its maximum? Do we have other possibilities such as the complexity may vibrate with different patterns before reaching its maximum? Intuitively the pattern of state complexity will be highly dependent on the state and the Hamiltonian. So maybe we have to admit that our universe is just one quantum system taking a specific initial state and a specific Hamiltonian among all the possible states and Hamiltonians. Different systems will have different state complexity patterns. and it's not unimaginable that there may exist a universe that the state complexity change of the black hole period is not perfectly linear so that before the system complexity reaches its maximum, the system complexity already shows a complex pattern. It's highly possible that our universe is also such a system. So in the black hole phase, the system will not visit all the low complexity states before it reaches the grey hole phase. Instead we may have multiple orbits or trajectories connecting an initial product state and a maximal complexity state. This gives a multiverse-like picture. Another observation is, if the possible multiple black hole states are all generated from a product state, then all these states can be connected by efficient quantum circuits so that the multiple spacetime of these different univeses can theoretically be connected. In another word, they all belong to the same quantum phase.

The near-maximal complexity grey phase will last for a double-exponentially long time, which is called the quantum recurrence time. What will be the pattern of the complexity fluctuation in this long recurrence time? We can imagine that it will be a zig-zag pattern, where the \emph{zig} denotes the complexity increasing phase and the \emph{zag} is the complexity decreasing phase. The problem is that how long the \emph{zig} and\emph{zag} can last. If they can last for a moderate long time, then our universe may also survive in this phase since we may not notice that we are actually in a zig/zag phase instead of a black/white hole phase. Can the high complexity states generate a similar world as the low complexity states in black hole phase? In \cite{Qi2013Exact} it's mentioned that in the holographic construction of spacetime, The bulk geometry is a holographic dual description of the boundary system. The bulk system can be regarded as a direct-product decomposition of the Hilbert space, in which entanglement between different sites is short-ranged even if in the original system the correlation and entanglement may be long-ranged. This seems a hint that the high complexity boundary states in the grey hole phase may also generate a low complexity bulk universes and zig-zag pattern seems to support a big bang/big bounce picture. Of course we have to admit, if Susskind is right, then the complexity decreasing zag phase will be very unstable and therefor can not last for a long time. Then the zig phase also has to have a short life. Then the complexity fluctuation in the grey hole phase is minor and we can not have a clear arrow of time, so that our universe can not stay there. Different with the black hole phase, the states in the grey hole phase can really belong to different quantum phases and there is no way to efficiently connect these states by efficient quantum circuits. If we regard this as another possibility of the multiverse picture, then it's likely that these multiple universes can not contact each other since there is no efficient quantum operation to build the spacetime to connect them.

The final white hole phase is not attractive to our discussion since it's highly unstable and we should not be in this phase.

The above discussion is based on the assumption that our observed universe is in a pure state. How about a mixed state universe as a subsytem of a much larger system? Do we still have the four phases? We assume at least the black/grey/white hole phases should still be there. The scrambler phase may still exist but the exponentially increasing pattern might need to be modified.

Besides all the above discussed uncertainties, we believe it's still safe to say, for a certain short enough period the monotonic state complexity pattern can be ensured and the arrow of time and the pace of time flow can be generated by the state complexity pattern, i.e., by the derivative of the complexity pattern. This can be regarded as the evidence that our hypotheses H1 and H2 are rational.

Hypothesis H3 has be partially testified in our former work \cite{Dong_time}, in which we show the arrow of time is closely related with the pattern of quantum state complexity.

\subsection{Quantum state and relativity}
Now we are in a position to check, given such a short-time stable complexity pattern, how we will interpret the complexity patterns and its related time flows when the quantum system is boosted or stays in a gravitational field.

This task steps into the field of relativistic quantum information, where related works showed that entanglement, purity, teleportation can all be Lorentz variant and acceleration dependent\cite{Peres2002Quantum}\cite{Timothy2012Relativistic}. Here we need to check, how will the quantum state complexity vary in a relativistic situation. We will examine this point using a simulation based approach.

Because our hypotheses H1 ahd H2 are built on the assumption of large many-body quantum systems, our simulation should be carried out on a large quantum system and we know this is very difficult if not impossible with classical computers. In this work we will focus on simulations on systems with only a few qubits to see if qualitatively our hypothesis H4 consists with relativity theory.

\textbf{Quantum state with special relativity}\\
 Given a quantum state $\rho$ of a quantum system, how will the state be understood by a boosted inertial observer? It's well known that there is a correspondence between Lorentz transformations and qubit operations. Following the idea of \cite{Arrighi2003A} we first explain how a boosted observer will interpret the state of a single qubit.

 A mixed state qubit can be mapped to a 4-vector as $\rho=\frac{1}{2}Tr(\rho\sigma_{\mu})\sigma_{\mu}$, where $\rho$ is the density matrix of the qubit and $\{\sigma_{\mu}\}_{\mu=0,...,3}$ designate the Pauli matrices $I,X,Y,Z$. The 4-vector is given by $\underline{\rho}=[\underline{\rho}_0,\underline{\rho}_1,\underline{\rho}_2,\underline{\rho}_3],Tr(\rho\sigma_{\mu})=Tr(\rho\sigma_{\mu})$.

 Then there is a mapping between the general quantum operation on $\rho$ and the Lorentz transformation on $\underline{\rho}$ as $\psi: SL(2,C)\rightarrow SO(1,3)^{+}$. With this correspondence, any Lorentz transform can be interpreted as a quantum operation on the qubit. Specially, the effect of a Lorentz boost on a qubit can essentially be viewed as applying a particular measurement clement $M$ whose outcome occurs.$M$ can be regarded as a nontrace-preserving quantum operation $MM^{+}\neq I$, which is in fact the $(\frac{1}{2},0)$ or $(0,\frac{1}{2})$ representations of a pure boost in Lorentz group. For example, a boost in x direction with the rapidity $\theta$ is given by $M=exp(X\theta/2)\in SL(2,C)$. Similarly a pure rotation corresponds to a unitary operation $U=exp(i\sigma_{\mu}\theta)$.

\textbf{Quantum state with general relativity}\\
For the case of general relativity, we consider the situation when the time metric is changed due to the existence of a gravitational potential. Following a similar idea of the special relativity case, a pure change on the time metric can be understood as a quantum operation given by $exp((I+X-Y+Z)\beta)$, where the factor $\beta$ controls the potential of the gravitational field.

Now we will check how the Lorentz transformation and the gravitational field will influence the quantum state complexity and therefore affect the flow of time.

\section{Quantum state complexity and special relativity}

\subsection{Quantum state complexity}
The quantum state complexity of a general N dimensional quantum state $\rho$ with a spectrum $\{\lambda_i\}_{i=1,...N}$ is defined as the minimal Bures distance between $\rho$ and all the states with a diagonal density matrix with the same spectrum of $\rho$. For example a single qubit $\rho$ with a spectrum ${\lambda_1,lambda_2}$, its complexity is $min(D_B(\rho,diag(\lambda_q,\lambda)2)),D_B(\rho,diag(\lambda_2,\lambda_1)))$ with $D_B(\rho_1,\rho_2)$ the Bures distance between states $\rho_1$ and $\rho_2$.

\subsection{Quantum state complexity of a 2 qubit system with special relativity}
As a warming up, we first check the quantum state complexity of the 2-qubit system discussed in \cite{Micadei_timearrow}\cite{Dong_time}. The 2-qubit system (A,B) has an initial state given by
\begin{equation}\label{eq_1}
  \rho_{AB}^{0}=\rho_{A}^{0}\otimes\rho_{B}^{0}+\chi_{AB},
\end{equation}
where $\chi_{AB}=\alpha|01\rangle\langle10|+\alpha^{*}|10\rangle\langle01|$ is the correlation term and $\rho_{i}^{0}=exp(=\beta_{i}H_{i})/z_{i}$ is a thermal state at inverse temperature $\beta_{i}$, $i=(A,B)$ for qubit A and B respectively.
The state $|0\rangle$ and $|1\rangle$ are the ground and the excited eigenstates of the Hamiltonian $H_{i}=h\nu_{0}(1-\sigma_z^{i})/2$. For simplicity, we set $h\nu_{0}=1$ so that $H_{i}=(1-\sigma_{z}^{i})/2$. Also the system will evolve under an effective interaction Hamiltonian $H_{AB}^{eff}=(\pi/2)(\sigma_{x}^{A}\sigma_{y}^{B}-\sigma_{y}^{A}\sigma_{x}^{B})$. In such a system, no work is performed and the heat absorbed by one qubit is given by its internal energy variation along the dynamics so that $Q_{i}=\Delta E_{i}$ with $E_{i}=Tr_{i}H_{i}\rho_{i}$.

In \cite{} we showed how the arrow of time is correlated with the change of the quantum state complexity. Here we will check how its state complexity will change under Lorentz transformations.
Following the above given mapping between Lorentz transformations and quantum information operations, we compute the state complexity evolving with the Hamiltonian $H_{AB}^{eff}$ as a transformation $exp(-iH_{AB}^{eff}\tau)$ with a latent time $\tau$ and then compute the state complexity under different pure rotations and pure boosts. The results are shown in \ref{}.

\begin{figure}
  \centering
  \includegraphics[width=9cm]{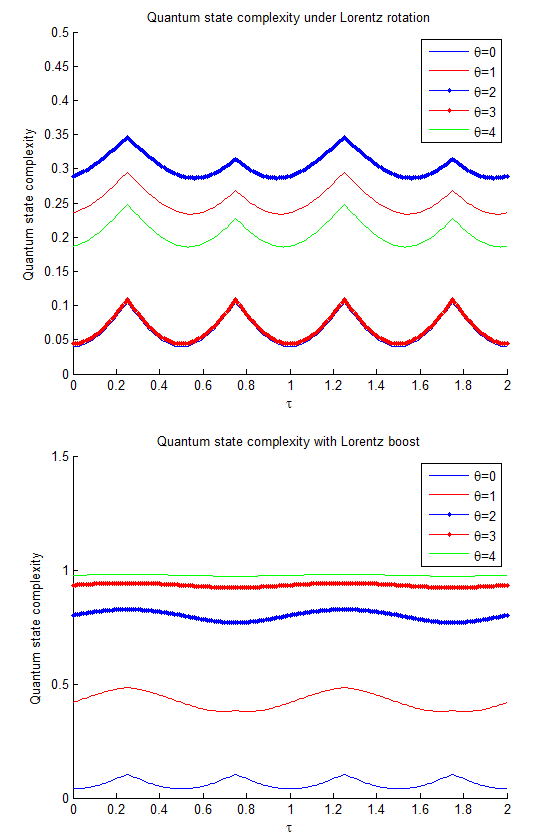}
  \caption{Patterns of quantum state complexity of an evolving 2-qubit system under Lorentz rotations $exp(-i(X_AI_B+I_AX_B)\theta/2)$ and boosts $exp((X_AI_B-I_AX_B)\theta/2)$. It can be observed that rotation and boost show different patterns, where rotations roughly keeps the derivative of state complexity but boosts trend to smear out the derivative.}\label{fig2}
\end{figure}

We can see that rotations keep the pattern of quantum state complexity but boosts smear the derivatives with an increasing rapidity. We believe this observation corresponds to the well known facts that rotations do not change the metric but boosts do. It need to be pointed out that in our simulation, the pattern of state complexity does change with rotations. This is due to the fact that we are working with a very simple 2-qubit system and the local unitary rotation operations can lead to a saturation of the state complexity. If we are working with a large many-body system with N qubits that is in the low complexity state as assumed by H1 and H2, the local unitary rotation operation can only change the state complexity linearly with a maximal change of complexity of $O(N)$.  So the state complexity pattern will be stable with rotations and rotations can uniformly shift the state complexity along the y axis in Fig. \ref{fig2}. Accordingly the derivative of state complexity with respect to $\tau$ will not be affected by rotations. On the contrary, boosts still change the state complexity dramatically with the rapidity becomes bigger and therefore the derivative of state complexity with respective to $\tau$ will be smeared out, which can be understood as the time dilation observed in special relativity. An easy way to understand the difference between rotations and boosts is that rotations do not change the spectrum of the quantum state and boosts change it. So the state complexity pattern is more stable under rotations than under boosts.

\subsection{Quantum state complexity of random states with special relativity}
Of course here we only checked one specific 2-qubit system with one specific Hamiltonian. To further testify our hypothesis H4, we carried out a simulational study on random quantum states as follows:
\begin{itemize}
  \item Generate a random low complexity N qubit system $\rho_0$ at the origin of an initial reference frame, where the complexity is computed according to our complexity definition.
  \item Choose a random local Hamiltonian $H$. The reason to choose local Hamiltonians is that we believe our universe obeys locality and it evolves under local operations.
  \item Evolve the system $\rho_0$ with $H$ for a short \emph{time} $\delta t$ to get $\rho_{\delta t}$.
  \item Compute the complexity of $\rho_0$ and $\rho_{\delta t}$ as $C(\rho_0)$ and $C_{\rho_{\delta t}}$. Then $v(\rho_0,H)=[C_{\rho_{\delta t}}-C_{\rho_{0}}]/\delta t$ is an approximation of the derivative of the state complexity. This is also the \emph{intrinsic} pace of time flow in this static reference frame.
  \item Compute the states and states complexity when the reference frame is under a pure rotation and a pure boost to get $\overline{\rho}_0, \overline{\rho}_{\delta t},C(\overline{\rho}_0), C(\overline{\rho}_{\delta t})$. Accordingly we can compute the pace of time flow under these Lorentz transformations as $\overline{v}(\rho_0,H)$.
  \item Repeat these steps for K times and compute the mean paces of time flows in different situations. Then we can compare the ratio of the paces of time flows between the Lorentz transformed case and the static reference case.
\end{itemize}

\begin{figure}
  \centering
  \includegraphics[width=10cm]{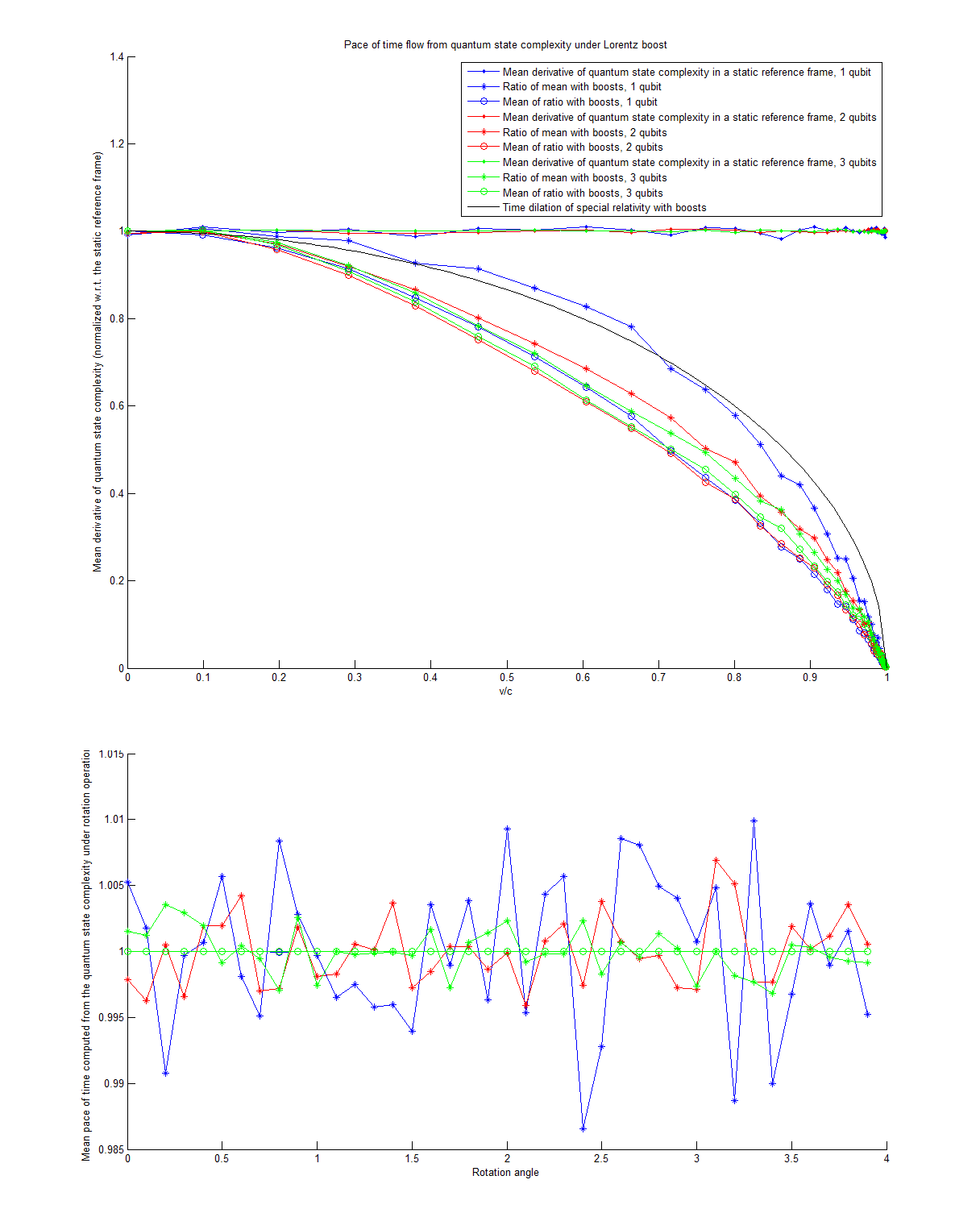}
  \caption{Simulation results of the pace of time flow as the derivative of quantum state complexity under rotations and Lorentz boosts. We simulated three simple systems, 1 qubit, 2 qubits and 3 qubits. Here we show the mean values of K=10000 examples. We see that rotations do not change pace of time flow and boosts will lead to a time dilation effect.}\label{fig3}
\end{figure}

We point out that there are two ways to compute the ratio of the paces of time flows with this random samples, either as \emph{the mean of ratio} given by $[\sum_{k=1}^K \frac{\overline{v}(\rho_0,H)^k}{v(\rho_0,H)^k}]/K$ or \emph{the ratio of mean} as $\frac{\sum_{k=1}^K\overline{v}(\rho_0,H)^k}{\sum_{k=1}^K v(\rho_0,H)^k}$. This is because we are not sure if our spacetime is a pure state or an ensemble of random history of spacetime. The simulation results for boosts and rotations are given in Fig. \ref{fig3}.

We can find a qualitative consistency between the changes of quantum state complexity under Lorentz transformations and special relativity. This is to say, a rotation does not change the time metric and a boost will lead to a time dilation. The reason that it's not a perfect match might be because we are not working with a large enough quantum system. Another reason is that in our simulation, the low complexity states are not the same as the states in the black hole phase. Because according to our definition, the zero complexity states with a diagonal density matrix can not be efficiently generated by a finite depth quantum circuit from a product state. Our simulated quantum states are more likely states in the grey hole phase and it's possible that the black hole phase and grey hole phase have different properties. Another reason may be that our quantum state complexity is computed on Bures distance and it may not be the proper metric for a qualitative evaluation of quantum state complexity to reproduce the time dilation effect of special relativity.

\section{Quantum state complexity and general relativity}

\begin{figure}
  \centering
  \includegraphics[width=9cm]{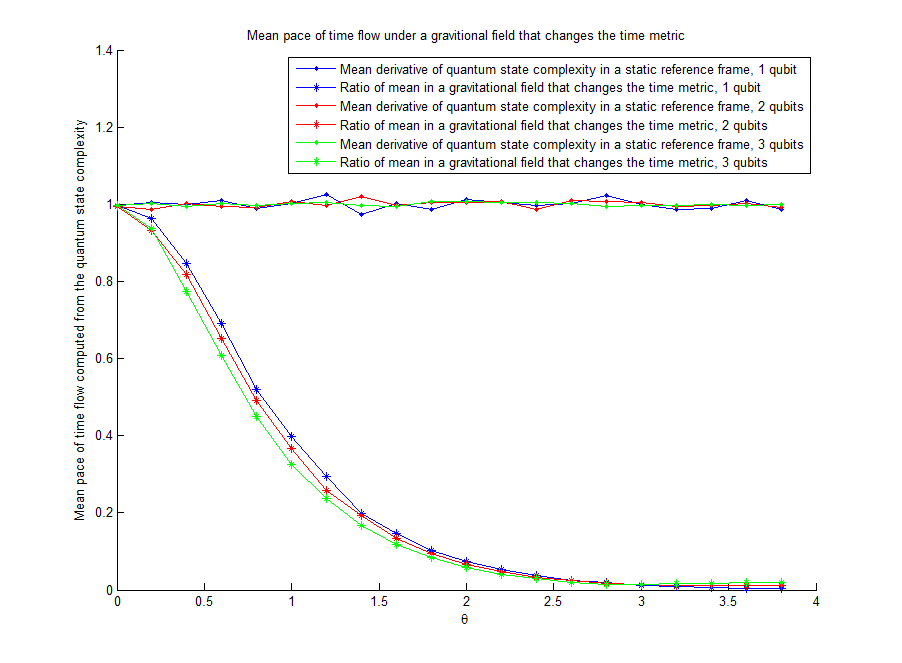}
  \caption{Simulation results of the pace of time flow based on quantum state complexity in a gravitational field which scales the time metric. We simulated three simple systems with 1, 2 and 3 qubits and the mean values are computed from 5000 random samples.}\label{fig4}
\end{figure}

We can also examine if our hypothesis H4 is valid when a gravitational field is present. Similar to the case of special relativity which keeps the metric invariant, the gravitational field that changes the metric can also be mapped to a quantum information operation. For example, with the $(\frac{1}{2},\frac{1}{2})$ representation, a scaling on the time metric while keeping the spatial metric invariant can be generated from $II+XX-YY+ZZ$ and $II+XX+YY-ZZ$,$-II+XX+YY+ZZ$,$II-XX-YY+ZZ$ generate a scaling of the spatial metric in x, y, z directions respectively. Following a similar idea in the special relativity to map a Lorentz operation to a quantum qubit operation, we can also map this time metric scaling transformation to a quantum information operation generated by $I+X-Y+Z$ on a single qubit. Then we can carry out a simulation of the quantum state complexity change under this transformation. The result on K=5000 random samples is shown in Fig. \ref{fig4}.

We see our simulation results show a time dilation which is computed from the quantum state complexity of different systems under time metric scaling transformations. Still we do not have a perfect reproduction of the time metric change due to the gravitational field. This can be due to an artifact of our model or too simple quantum sysmtems or the Bures distance. But at least quantum state complexity can lead to a phenomenon that looks like the time dilation in this case.

\section{Quantum state complexity and thermal time hypothesis}
Finally we will check the relationship between our time from quantum state complexity idea and the thermal time hypothesis (TTH)\cite{Connes1994Von}\cite{Rovelli2010Thermal}\cite{Menicucci2011Clocks}. TTH aims to find out the origin of the time flow in a generally covariant situation. TTH claims: The physical time depends on the state. When the system is in a state $\omega$, the physical time is given by the modular group $\alpha_t$ of $\omega$ and this time is called the thermal time.

TTH assets that any arbitrary state $\rho$ defines a thermal time given by the flow of $H_{\rho}=-ln(\rho)$ on the observables, with respect to which the state $\rho$ appears thermal. It has been shown to agree with known physics in different contexts such as the Unruh temperature and Tolman-Ehuenfest effect\cite{Connes1994Von}\cite{Rovelli2010Thermal}. In \cite{Cao2017Space} a geometry from bulk entanglement scheme was proposed to show that entanglement perturbations of a state give rise to a local modification of spatial curvature, which is an spatial analog of Einstein's equation. There the modular Hamiltonian, which is equivalent to $H_{\rho}$, is used to derive the relationship between modular energy and the change and entropy of a state, which further leads to a change of the curvature of the emergent space, i.e. the gravity. We see here that similarly a perturbation of the state dependent $H_{\rho}$ can also lead to a change of the state complexity and thereby results in a change of the time metric.

Now we show that our idea of time from quantum state complexity can be regarded as a generalization of TTH. We will see the time from quantum state complexity idea on one hand claims that the time flow is the same as the quantum state complexity flow and therefore time is state dependent. On the other hand, the quantum state complexity as a Riemannian metric reveals the statistical property of time due to the geometric picture of quantum mechanics. These are exactly the key ideas of TTH.

In our hypothesis H1-H3, we assume that our universe is in in a low complexity state that is generated by a local quantum circuit from a simple product state. This is to say, our universe as a huge quantum system, its state evolves as $i\frac{\partial}{\partial t}\rho(t)=[\rho(t),H(t)]$ with $H(t)$ a local Hamiltonian and $\rho(0)$ a rank-1 diagonal matrix. Easily we have $\rho(t)=U(t)\rho(0)U(t)^+$ with $U(t)=exp(-i\int_0^t H(s)ds)$. The state complexity of $\rho(t)$ is given by the length of the minimal quantum circuit achieving the unitary operation $U(t)$ as in Nielsen's quantum computation complexity theory, which is equivalent to the length of the geodesic connecting $\rho(0)$ and $\rho(t)$ in the quantum state space. Besides this Riemannian manifold picture, we also have a Lie group picture of quantum state complexity, in which the unitary operation space is regarded as a Lie group and any unitary operation $U$ is generated by a Lie exponential as $U=exp(-iH)$ with $H$ a constant Hamiltonian. Then the complexity of $\rho(t)=U(t)\rho(0)U(t)^+$ is given by the length of a vector in its correspondent Lie algebra $ln(U(t))=-iH$, which is equivalent with the operator norm $\|H\|$. for a large quantum state, even in the Riemannian manifold picture where the quantum state complexity is essentially the geodesic connecting identity operator $I$ and $U(t)$, the state complexity can still be approximated by $\|H\|$ with $U(t)=exp(-iH)$ or by $\int_s \|H(s)\|ds$ with $U(t)=exp(-i\int_0^t H(s)ds)$. This picture can be generalized to mixed state so that a general quantum state $\rho(t)$ can be written as $\rho(t)=exp(-i\int H(s)ds)\rho(0)exp(-i\int H(s)ds)$ with $\rho(0)$ a diagonal density matrix, which represents a zero complexity state of our definition of the complexity of a quantum state.

Now it's easy to see that TTH is just a special case of our time from quantum state complexity picture when $H(t)$ is constant and $\rho(0)=I$. In this case, we have $\rho(t)=exp(-i2Ht)$ so that $-ln\rho(t)=i2Ht$ and the state complexity of $\rho(t)$ is given by $\|H\|t$.  Obviously there exists a mapping between $\rho(t)$ and $U(t)=exp(-iHt)$. If we modify the $-ln(\rho)$ of TTH as $-ln(U(t))=iHt$, comparing this with TTH, we immediately see the flow of the quantum state complexity $\rho(t)$ corresponds to the time flow of TTH.

Given this relationship, a quantum state $\rho(t)=U(t)\rho(0)U(t)^+$ with $U(t)=exp(-iHt)$, the TTH thermal time is related with the mechanical time by a factor $t=1/kT$ with $T$ the temperature\cite{Rovelli2010Thermal}. We immediately see $\beta=1/T\propto t$ is proportional to the quantum state complexity of $\rho(t)$ given by $\|H\|t$ approximately. So applying TTH to such a system, a higher complexity corresponds to a lower temperature. This is consistent with our hypothesis that our universe evolves from a low complexity to a higher complexity so that the temperature of our universe decreases with time.

Another observation is that in TTH, if we partitioning the complete system into subsystems, then each system has its own thermal time and all these local clocks need to be synchronized in a natural way. In our time from quantum complexity picture, the synchronization of local clocks can be easily achieved if $H(t)$ is a local Hamiltonian and $\|H(t)\|$ is uniformly distributed in all subsystems. This means the state complexities of all subsystems evolve with roughly the same speed with respect to their system sizes. In our picture, this means the pace of time is constant in all subsystems.

\section{Conclusion}
In this work we developed the hypothesis of time from quantum state complexity. We proposed that the concept of time origins from the change of quantum state complexity. The direction of the thermodynamic arrow of time points in the direction of increasing quantum state complexity, and the pace of time flow is determined by the changing rate of quantum state complexity. We then verified this hypothesis on simple qubit based quantum systems to show that this hypothesis is consistent with the time dilation effect of both special and general relativity theories. We also examined the relationship between our idea and the thermal time hypothesis to show that they are also consistent.

We also discussed the quantum state of our universe. The reason that we have a smooth time flow lies in that our universe is in a low complexity quantum state starting from a simple product state, whose complexity increases linearly. We discussed the possibility that the pace of time of our universe might be time variant depending on the state and Hamiltonian of our universe. We also considered the possibility that our universe is in the grey hole phase so that this may support a big bang/big bounce picture and a multiverse picture as well. We suspected that this quantum state complexity perspective may be related with the ideas of big bang, inflation and multiverse.

\bibliographystyle{unsrt}

\bibliography{timearrow}





\end{document}